\pdfoutput=1
\documentclass{ifacconf}

\counterwithin*{section}{part}

\usepackage[printonlyused,withpage]{acronym} % withpage viser hvilke sider akronymer er brukt når de til slutt printes, printonlyused sørger for at kun definerte akronymer som faktisk er brukt i teksten vil printes i akronymoversikten
\usepackage[acronym]{glossaries-extra} % no greier
\setabbreviationstyle[acronym]{long-short}
\usepackage{graphicx}      % include this line if your document contains figures
\usepackage{natbib}        % required for bibliography
\usepackage{amsmath, amssymb}
\usepackage{bm}
\usepackage{siunitx}
\usepackage{mathtools}
\usepackage{setspace}
\usepackage{listings}
\usepackage[capitalise]{cleveref}
\usepackage{etoolbox}
\crefname{equation}{}{}
\crefformat{footnote}{#2\footnotemark[#1]#3}
\newcommand\vm[1]{% Vector or matrix
\bm{\mathrm{#1}}}

\newcommand\derp[1]{\frac{\partial}{\partial#1}}

\GlsXtrEnableEntryUnitCounting{acronym}{1}{document}

\begin{document}
\setlength{\belowdisplayskip}{1pt}
\setlength{\abovedisplayskip}{1pt}
\setlength{\intextsep}{3pt}
\setlength{\dblfloatsep}{0pt}
\setlength{\floatsep}{0pt}
\setlength{\abovecaptionskip}{0pt}
\setlength{\belowcaptionskip}{0pt}
\setlength{\textfloatsep}{5pt}
%\belowcaptionskip{0pt}
%\abovecaptionskip{-100pt}
\linepenalty=3500

\newacronym{hzd}{HZD}{hybrid zero-dynamics}
\newacronym{cot}{CoT}{cost of transport}
\newacronym{rms}{RMS}{root mean square}
\newacronym{mpc}{MPC}{model predictive control}
\newacronym{nmpc}{NMPC}{nonlinear model predictive control}
\newacronym{clf}{CLF}{control-Lyapunov function}
\newacronym{lfc}{LFC}{Lyapunov function candidate}
\newacronym{nlp}{NLP}{nonlinear programming}
\newacronym[longplural={bilinear matrix inequalities}]{bmi}{BMI}{bilinear matrix inequality}
\newacronym[longplural={linear matrix inequalities}]{lmi}{LMI}{linear matrix inequality}
\newacronym{rl}{RL}{reinforcement learning}
\newacronym[longplural = {hybrid dynamical systems}, shortplural = {HDS}]{hds}{HDS}{hybrid dynamical system}
\newacronym{zmp}{ZMP}{zero moment point}
\newacronym{cog}{CoG}{center of gravity}
\newacronym{srbd}{SRBD}{single rigid-body dynamics}
\newacronym{ltv}{LTV}{linear time-varying}
\newacronym[longplural={linear independence constraint qualifications}]{licq}{LICQ}{linear independence constraint qualification}
\newacronym{nloc}{NLOC}{nonlinear optimal control}
\newacronym{lq}{LQ}{linear-quadratic}
\newacronym{ad}{AD}{automatic differentiation}
\newacronym{pe}{PE}{persistently exciting}
\newacronym{drl}{DRL}{deep reinforcement learning}
\newacronym[longplural={degrees of freedom}, shortplural={DOFs}]{dof}{DOF}{degree of freedom}
\newacronym{dh}{DH}{Denavit-Hartenberg}
\newacronym[longplural={centers of mass}, shortplural={CoMs}]{com}{CoM}{center of mass}
\newacronym{ode}{ODE}{ordinary differential equation}
\newacronym{phzd}{PHZD}{partial hybrid zero dynamics}
\newacronym{mpp}{MPP}{Moore-Penrose pseudoinverse}
\newacronym{urdf}{URDF}{unified robot description format}
\newacronym{eos}{EOS}{exponentially orbitally stable}
\newacronym{qp}{QP}{quadratic programming}
\newacronym{wbc}{WBC}{whole body controller}

\begin{frontmatter}

\title{Torque-Minimizing Control Allocation for Overactuated Quadrupedal Locomotion}

\thanks[footnoteinfo]{This result is part of a project that has received funding from the European Research Council (ERC) under the European Union’s Horizon 2020 research and innovation programme, through the ERC Advanced Grant 101017697-CRÈME. The work is also supported by the Research Council of Norway through the Centres of Excellence funding scheme, project No. 223254 – NTNU AMOS.}

\author[First]{Mads Erlend Bøe Lysø} 
\author[Second]{Esten Ingar Grøtli}
\author[First]{Kristin Ytterstad Pettersen} 

\address[First]{Department of Engineering Cybernetics, Norwegian University of Science and Technology (NTNU), Trondheim, Norway (e-mail: \{mads.e.b.lyso,kristin.y.pettersen\}@ntnu.no)}
\address[Second]{Department of Mathematics and Cybernetics, SINTEF Digital, Trondheim, Norway (e-mail: EstenIngar.Grotli@sintef.no)}

\begin{abstract}
\label{sec:abstract}
In this paper, we improve upon a method for optimal control of quadrupedal robots which utilizes a full-order model of the system. The original method utilizes offline nonlinear optimal control to synthesize a control scheme which exponentially orbitally stabilizes the closed-loop system. However, it is not able to handle the overactuated phases which frequently occur during quadrupedal locomotion as a result of the multi-contact nature of the system. We propose a modified method, which handles overactuated gait phases in a way that utilizes the full range of available actuators to minimize torque expenditure without requiring output trajectories to be modified. It is shown that the system under the proposed controller exhibits the same properties, i.e. exponential orbital stability, with the same or lower point-wise torque magnitude. A simulation study demonstrates that the reduction in torque may in certain cases be substantial.
\end{abstract}

\begin{keyword}
Robotics, Non-Linear Control Systems, Control Design, Optimal Control, Discrete Event and Hybrid Systems
\end{keyword}

\end{frontmatter}

\section{Introduction}
\vspace{-3mm}
% Hvorfor er denne forskningen relevant/viktig?
Quadrupedal robots have features that are beneficial for achieving dynamically stable and robust walking: Legged robots in general are able to adapt well to rough terrain while keeping the main base levelled, and the multi-contact nature of quadrupeds lend then some inherent robustness not shared by bipedal robots. However, the multi-contact interaction with its environment, combined with the high dimensionality of such systems, makes the modeling and control of such robots challenging. This has led to a greater reliance on clever heuristics and lower-order models when developing methods for control. As a result, progress has been made predominantly in simulations and in real-world experiments without a corresponding mathematical analysis or proof of theoretical stability. However, in recent years we have seen both increases in computational power and a greater understanding of and ability to exploit the structure of the control problem. As a result, it is increasingly feasible to utilize the full kinematic, and in some cases dynamical, models of the robot in offline and online methods for control. However, the constrained nature of these systems often leads to phases of the gait in which the ground contacts render the system dynamics overactuated. This complicates the control problem by making the series of inputs that correspond to a desired system behavior non-unique.
%\begin{figure}
%    \centering
%    \includegraphics[trim=50 200 650 1000,clip,width=0.48\textwidth]{figures/astronomical.png}
%    \caption{Photo of the sprawling quadrupedal robot ASTRo}
%    \label{fig:astro}
%\end{figure}
\begin{figure}
    \centering
    \includegraphics[width=0.48\textwidth]{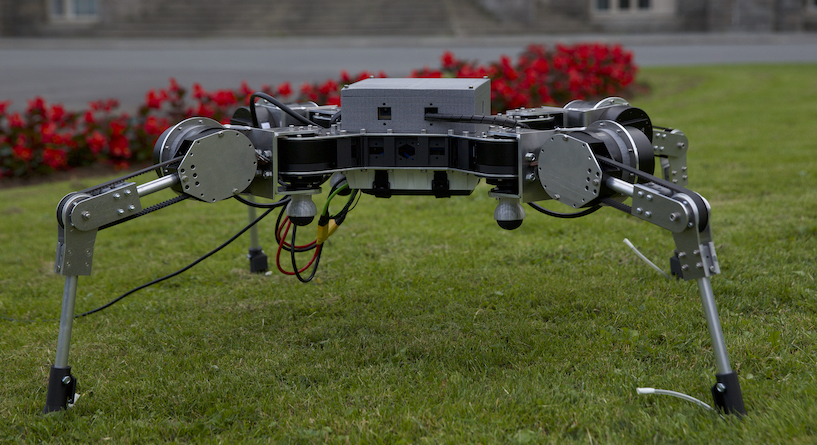}
    \caption{Photo of the sprawling quadrupedal robot ASTRo}
    \label{fig:astro}
\end{figure}

%\subsection{Related work}
Several approaches have been taken in recent years to incorporate more detailed knowledge of the system dynamics into the control of quadrupedal robots. One popular avenue through which this has been explored is the control paradigm of \gls{mpc}. While certain methods such as \cite{di_carlo_dynamic_2018} have opted for fairly reduced-order models in order to yield linear \gls{mpc} control schemes, others have attempted to more fully capture the complexities of the system dynamics. \cite{neunert_whole-body_2018} utilize a full-order nonlinear model of the robotic system to produce a \gls{nmpc} scheme. One of the advantages of this method is that the choice of gait pattern is also subject to online optimization, which is achieved through using a smooth approximation of the ground contact forces. In \cite{minniti_adaptive_2022}, a system model using the full kinematics and the dynamics of the base is utilized in an adaptive \gls{nmpc} scheme. The scheme produces a state-feedback controller that asymptotically stabilizes the system to a desired trajectory in the presence of modeling errors. A \gls{qp}-based \gls{wbc} is used to calculate the joint torques to track the desired accelerations and contact forces.

Another approach is to incorporate a high-fidelity robot model into an offline optimization problem. An example of this approach is presented in \cite{ma_first_2019}. Here, the optimization problem describes a closed-loop system subject to a parametrized controller. The result of the optimization is a set of parameters define a closed-loop controller. Thus, no auxiliary control frameworks such as a \gls{wbc} are needed to track the returned optimal trajectory. One benefit is that in contrast to e.g. a \gls{wbc}-scheme, which may often require solving an optimization problem, the resulting controller here often has a much simpler structure. Specifically, the controller incorporated in \cite{ma_first_2019} is an input-output linearization controller, which exhibits properties such as Lipschitz continuity and continuous differentiability that may be important for stability analysis. The closed-loop system behavior is shown to be \gls{eos}.

Finally, there have been some efforts in the later years to capture complex dynamics and scenarios without resorting to high-fidelity explicit modeling. Notably, in \cite{miki_learning_2022}, a method is presented utilizing \gls{drl} to produce an end-to-end model-free locomotion controller. The controller learns the most efficient trade-off between exteroception and proprioception for a wide range of scenarios. The empirical results of the paper show the robot traversing a mountain hiking route to the top faster than the nominal tourist traversal time.

While several of the above-mentioned methods yield good empirical results, most of them do not have theoretical guarantees for performance in the form of stability proofs. Even in \cite{minniti_adaptive_2022}, where a proof of convergence to a nominal trajectory is shown, this holds only for the simplified model. Also, stable trajectories for the span of the \gls{mpc} horizon do not guarantee the generation of feasible, stable trajectories in the long run. In this respect, the controller presented in \cite{ma_first_2019} seems to have a clear advantage: The synthesized controller stabilizes the full dynamics exponentially to a submanifold of the state space, which again contains an \gls{eos} orbit.

However, the method as described does not cope satisfactorily with the following challenge: During some phases of a gait cycle, the robot has two legs in contact with the ground. During these phases the robot is over-actuated, and there is an infinite set of inputs that will result in exactly the same acceleration. \cite{ma_first_2019} handles this problem by disabling one actuator in such cases, returning the system to a state of non-overactuation. This is not necessarily a satisfying solution. Firstly, the choice of which actuator to disable is not explicitly justified, and it is not clear what method one would use to make this decision. Secondly, the analogy would be for a human standing on two legs to disable its thigh muscles to avoid overactuation, which is not obviously the most robust nor energy-efficient answer to the question of torque-allocation. One way to handle this issue could be to simply opt for e.g. a \gls{qp}-based \gls{wbc}, which is capable of choosing from this set of possible inputs subject to some objective, e.g. minimization of contact forces or torques. However, the simple, transparent structure of a closed-form controller would then be lost. Moreover, properties such as Lipschitz continuity are not immediately present in such kinds of controllers, which may interfere with the stability analysis.

%\subsection{Contribution}
In this paper, we modify the method presented in \cite{ma_first_2019} to achieve point-wise optimal (in a least-squares sense) control allocation for a given desired output trajectory. Specifically, we alter the control structure to allow for a greater number of actuators than actuated \glspl{dof}. Thus, instead of removing the resulting ambiguity in the choice of inputs by disabling the superfluous actuators, our controller utilizes this ambiguity to choose the smallest of the inputs which realize the output trajectory. As in \cite{ma_first_2019}, the resulting closed-loop system behavior exhibits exponential stability of the outputs, and exponential orbital stability of the full system state modulo horizontal position. Furthermore, the resultant torque expenditure is point-wise less than or equal to the torque expenditure of the previous method, without requiring any modification of the system trajectories. Following this, the controller is implemented on a model of the sprawling quadrupedal robot ASTRo, which was designed and built in \cite{ghansah_design_2021}\footnote{\label{note1}By a mistake, this reference was not included in the article \cite{lyso_torque_minimizing_2023} published in the 2023 IFAC WC proceedings.}, in a simulation study. The study demonstrates that the torque reduction is fairly substantial in certain cases, as is the reduction in energy expenditure that follows.

%\subsection{Outline}
The paper is organized as follows. In \cref{sec:system_modeling} we briefly describe the dynamic system formulation. In \cref{sec:methods} we first describe the optimal control formulation and the time-varying input-output linearization scheme as is used in \cite{ma_first_2019}, as well as the method for exponential stabilization, which we build upon. Then, we present the modified controller which readily handles overactuated phases of the gait. We then present simulation results illustrating the proposed method and showing the stability and energy-expenditure in \cref{sec:results}. Finally, in \cref{sec:conclusion} we present the conclusions and discuss future work.
\vspace{-2.7mm}
\section{System modeling}
\label{sec:system_modeling}
\vspace{-3mm}
In this section, we briefly describe the model of the system. The robotic system to be modeled is the quadrupedal robot ASTRo, short for Articulated Sprawling Tetrapod Robot \citep{ghansah_design_2021}\cref{note1}, see \cref{fig:astro}. The robot has a floating base which is not directly actuated, as well as four legs, each with three actuated joints. Each leg first has a hip link connected to the base by a joint whose rotational axis is normal to the dorsal plane. Then, the upper and lower leg links are connected in series with joints whose rotational axes lie in the dorsal plane of the robot. The base has 3 positional and 3 orientational \glspl{dof}, with the orientation parametrized by Tait-Bryan angles, while each leg has 3 joints. In total, this results in 18 \glspl{dof} for the system. The generalized coordinates become $\vm{q} = \left(\vm{p}_b^{\top}, \vm{\phi}_b^{\top}, \vm{\theta}_{\textnormal{legs}}^{\top}\right)^{\top}$, where $\vm{p}_b$ and $\vm{\phi}_b$ represent the position and orientation of the base, while $\vm{\theta}_{\textnormal{legs}}$ represents the joint angles of the legs. The following transcription of the ASTRo robot into a \gls{hds} is based on work found in \cite{hereid_dynamic_2018,ma_first_2019}, but there performed respectively for a bipedal and mammalian quadrupedal robot.
\vspace{-2mm}
\subsection{Hybrid Dynamical System Formulation}
\vspace{-2mm}
A walking robot can be described as a \gls{hds}. Adhering to the formulation in \cite{hamed_dynamically_2019}, the \gls{hds} can be described by a set of continuous system dynamics $FG = \left\{(f_v, g_v)\right\}$ on a set of continuous domains $\mathcal{D} = \left\{\mathcal{D}_v\right\}$ so that $\dot{\vm{x}}_v = f_v(\vm{x}_v) + g_v(\vm{x}_v)\vm{u}_v$ where $\vm{x}_v$ is the state and $\vm{u}_v$ is the input, and $(\vm{x}_v, \vm{u}_v)\in\mathcal{D}_v$. Each domain and system of equations correspond to a vertex $v\in\mathcal{V}$ of a graph $\Lambda = (\mathcal{V}, \mathcal{E})$, with $\mathcal{E}$ being the set of edges. In our case we denote the state $\vm{x}_v = (\vm{q}_v^{\top}, \dot{\vm{q}}_v^{\top})^{\top}$. We define the successor function $\mu:\mathcal{V}\rightarrow\mathcal{V}$ so that $v_2 = \mu(v_1)$ if there is an edge $(v_1\rightarrow v_2) \in \mathcal{E}$. Transitions between domains are governed by a set of guards $\mathcal{S}=\left\{\mathcal{S}_{(v\rightarrow\mu(v))}\right\}_{(v\rightarrow\mu(v))\in\mathcal{E}}$ and discrete dynamics $\Delta = \left\{\Delta_{(v\rightarrow\mu(v))}\right\}_{(v\rightarrow\mu(v))\in\mathcal{E}}$. Each guard is a surface in $\mathcal{D}_v$ so that a transition occurs from $v$ to $\mu(v)$ when $\vm{x}_v\in\mathcal{S}_{(v\rightarrow\mu(v))}$. The transition occurs according to $\vm{x}_{\mu(v)} = \Delta_{(v\rightarrow\mu(v))}(\vm{x}_v)$.
\vspace{-2mm}
\subsection{Continuous dynamics and constraints}
\vspace{-2mm}
We introduce the index set $\mathcal{I}_{\textnormal{legs}} = \left\{\textnormal{fl}, \textnormal{rl}, \textnormal{fr}, \textnormal{rr}\right\}$ and refer to the vector of joint angles for leg $i$ as $\vm{\theta}_i$, e.g. $\vm{\theta}_{\textnormal{fl}}$ for the front left leg. Furthermore, the robot has some of its legs in contact with the ground at any given time, encoded by the index set $\mathcal{I}_c \subseteq \mathcal{I}_{\textnormal{legs}}$. We here model the feet as point feet, and assume a positional no-slip condition and zero rotational friction. The first is a standard assumption in the literature, the second is a modeling choice. In \cite{hereid_dynamic_2018} a rotational no-slip condition on the surface normal axis is enforced as well. However, the frequent multi-contact phases and the sprawling configuration is likely to make such a condition too restrictive for our robot due to the alignment between the hip joint rotational axis and the surface normal, which is not present in the mammalian configuration. The no-slip conditions are described by a vector of kinematic constraints
\begin{equation}
    \label{eq:kinematic_constraints}
    \vm{g}_c(\vm{q}) = \textnormal{vec}\left(\left\{\vm{p}_i(\vm{q}) - \vm{p}_{i}(\vm{q}(0)) = \vm{0}\right\}_{i \in \mathcal{I}_c}\right)
\end{equation}
with $\vm{p}_i(\vm{q})$ denoting the position of the point-foot $i$ in the world frame. From the Euler-Lagrange equation we get our equations for the constrained nonlinear dynamical system:
\begin{subequations}
\begin{align}
    \vm{D}(\vm{q})\ddot{\vm{q}} + \vm{C}(\vm{q}, \dot{\vm{q}})\dot{\vm{q}} + \vm{G}(\vm{q}) &= \vm{B}\vm{u} + \vm{J}_c^{\top}(\vm{q})\vm{\lambda} \label{eq:system_dynamics_1}\\
    \vm{J}_c(\vm{q})\ddot{\vm{q}} + \derp{\vm{q}}\left(\vm{J}_c(\vm{q})\dot{\vm{q}}\right) &= 0 \label{eq:system_dynamics_2}
\end{align}
\label{eq:system_dynamics}%
\end{subequations}
where $\vm{B} = \left[\vm{0}_{6\times12}^{\top}, \vm{I}_{12}\right]^{\top}$ is the actuation matrix, $\vm{D}(\vm{q})$ is the mass matrix, $\vm{C}(\vm{q},\dot{\vm{q}})$ is the Coriolis matrix, and $\vm{G}(\vm{q})$ represents the potential terms. The Lagrange multipliers of the constraint forces are denoted by $\vm{\lambda}$, while $\vm{J}_c(\vm{q}) = \derp{\vm{q}}\vm{g}_c(\vm{q})$. We may re-arrange these equations to the explicit, control-affine form assumed in the \gls{hds} formulation:
\begin{equation}
    \label{eq:nonlinear_system}
    \dot{\vm{x}}_v = \vm{f}_v(\vm{x}_v) + \vm{g}_v(\vm{x}_v)\vm{u_v}
\end{equation}
\vspace{-2mm}
\subsection{Discrete dynamics and transitions}
\vspace{-2mm}
Each discrete transition between subsystems represents either the landing or lift-off of a foot. For transitions that correspond to lift-offs, the discrete dynamics are simply the identity. For transitions corresponding to feet hitting the ground, however, we assume a perfectly plastic impact; this is a typical choice in the literature to avoid the identification of parameters related to elastic impact \cite[p.~12]{grizzle_models_2014}. This implies that there is no abrupt change in $\vm{q}$, but the velocity of the impact foot immediately becomes $0$. From this and the conservation of momentum, we get the discrete dynamics
\begin{equation}\label{eq:discrete_dynamics}
    \Delta_{(v\rightarrow\mu(v))}(\vm{x}_v) =
    \begin{pmatrix}
        \vm{q}_v\\
        \dot{\vm{q}}_v + \vm{D}^{-1}(\vm{q}_v)\vm{J}_{\mu(v)}(\vm{q}_v)\vm{\lambda}_{\textnormal{impulse}}
    \end{pmatrix}
\end{equation}
%\begin{equation}\label{eq:discrete_dynamics}
%\begin{split}
%    \Delta_{(v\rightarrow\mu(v))}(\vm{x}_v) =
%    \begin{cases}
%    \begin{pmatrix}
%        \vm{q}_v\\
%        \dot{\vm{q}}_v + \vm{D}^{-1}(\vm{q}_v)\vm{J}_{\mu(v)}(\vm{q}_v)\vm{\lambda}_{\textnormal{impulse}}
%    \end{pmatrix}
%    \\%&\textnormal{ for impact}\\
%    \vm{x}_v% &\textnormal{ for lift-off}
%    \end{cases}
%\end{split}
%\end{equation}
for impact, where $\vm{\lambda}_{\textnormal{impulse}}$ is the intensity of the contact impulse required to set all foot velocities to $0$. For lift-off, the dynamics simply become $\Delta_{(v\rightarrow\mu(v))}(\vm{x}_v) = \vm{x}_v$.
\vspace{-2mm}
\subsection{Guards and admissible domain}
\vspace{-2mm}
For each vertex, the admissible domain $\mathcal{D}$ is described as the states satisfying a set of constraints. Firstly, there are constraints relating to the contact forces, denoted $\vm{v}_v(\vm{q}_v, \dot{\vm{q}}_v)\vm{\lambda}_v(\vm{q}_v) \geq \vm{0}$. Specifically, the contact force must not exit the linearized friction cone so as to not slip. Secondly, there are constraints independent of the contact forces. We denote these as $\vm{h}_v(\vm{q}_v, \dot{\vm{q}}_v) \geq \vm{0}$. Here, we require the position of all swing feet to be greater than 0, i.e., above the ground. These constraints are summarized as
\begin{equation}
    \vm{A}_v =
    \begin{bmatrix}
        \vm{v}_v(\vm{q}_v, \dot{\vm{q}}_v)\vm{\lambda}_v(\vm{q}_v)\\
        \vm{h}_v(\vm{q}_v, \dot{\vm{q}}_v)
    \end{bmatrix}
    \geq \vm{0}
    \label{eq:admissible_domain_vector}
\end{equation}

Each guard is the subset of the boundary of the domain, and a transition occurs when the state is about to exit the domain through it. Thus, for a select element $H_{(v\rightarrow\mu(v))}$ from \eqref{eq:admissible_domain_vector}, we may define a corresponding guard as
\begin{equation}
    \mathcal{S}_{(v\rightarrow\mu(v))} =
    \left\{
    (\vm{q}_v, \dot{\vm{q}}_v, \vm{u}_v) \middle\vert
    \begin{array}{c}
        H_{(v\rightarrow\mu(v))}(\vm{q}_v, \dot{\vm{q}}_v) = 0\\
        \dot{H}_{(v\rightarrow\mu(v))}(\vm{q}^{-}_v,\dot{\vm{q}}^{-}_v) < 0
    \end{array}
    \right\}
\end{equation}
%\begin{equation}
%    \mathcal{S}_{v\rightarrow\mu(v)} = \{(\vm{q}, \dot{\vm{q}}, \vm{u}) \: | \: H_{v\rightarrow\mu(v)}(\vm{q}_v, \dot{\vm{q}}_v) = 0, \: \dot{H}_{v\rightarrow\mu(v)}((\vm{q}^{-}_v, \dot{\vm{q}}^{-}_v)) < 0\}
%\end{equation}
with $\dot{H}_{(v\rightarrow\mu(v))}(\vm{q}^{-}_v,\dot{\vm{q}}^{-}_v)$ denoting the left hand derivative with respect to time. For transitions corresponding to a foot lift-off, $H_{(v\rightarrow\mu(v))}$ is the normal component of the contact force of the lift-off foot. For transitions corresponding to an impact, $H_{(v\rightarrow\mu(v))}$ is the impact foot height.
\vspace{-3mm}
\section{Methods}
\label{sec:methods}
\vspace{-3mm}
In this section we briefly describe the methods we build on from \cite{ma_first_2019}, before we introduce our suggested modification to handle overactuated phases.
\vspace{-2mm}
\subsection{Offline optimal control}
\vspace{-2mm}
\label{sec:optimal_control}
As in \cite{ma_first_2019} we use the Fast Robotics Optimization and Simulation Toolkit (FROST), first presented in \cite{hereid_frost_2017}, to perform offline optimal control of the \gls{hds} described in \cref{sec:system_modeling}. For the sake of the optimization problem each domain has only one possible successor, and the domains in this cycle determines the gait pattern. The duration of each phase is subject to optimization. Periodicity is enforced by the boundary condition
\begin{equation}
    \label{eq:boundary_condition}
    \vm{x}(t_0) + (t_f - t_0)\vm{v}_{\textnormal{des,xy}} - \Delta_{v_f\rightarrow\mu(v_f)}(\vm{x}(t_f)) = \vm{0}
\end{equation}
where $v_f$ is the final vertex of $\Lambda$, and $\vm{v}_{\textnormal{des,xy}}$ is nonzero in the first two elements and describe the desired horizontal velocity. $t_0$ and $t_f$ denote the initial and final times.

FROST utilizes a direct collocation transcription scheme and pre-compiled symbolic representations of constraints to make optimization over high-dimensional and highly complex systems computationally feasible. The framework is designed to incorporate control laws into the optimization problem itself, so that solution contains a set of controller parameters $\vm{\alpha}^{*}$ which define a control law $\vm{u}(t, \vm{x}, \vm{\alpha}^{*})$. Please see \cite{hereid_dynamic_2018} for a detailed description of FROST. The objective function is in this case chosen as the sum of the squared 2-norm of torque expenditure, divided by the period of the resulting gait. The division is done so as to avoid the solver favoring shorter gaits.

The discretized optimization problem then becomes
\begin{equation}
\begin{split}
    &\min_{\left\{\vm{x}_k\right\}, \left\{\vm{u}_k\right\}, \left\{\vm{\alpha}_k\right\}, \left\{t_k\right\}} \frac{1}{t_f - t_0}\sum_{k} \vm{u}_k^{\top}\vm{u}_k\\
    &\textnormal{s.t.}\\
    &\textnormal{System dynamics as per \cref{eq:system_dynamics,eq:discrete_dynamics}(transcribed)}\\
    &\textnormal{Chosen control law } \vm{u}(t,\vm{x},\vm{\alpha})\\
    &\textnormal{State constraints including \eqref{eq:admissible_domain_vector}}\\
    &\textnormal{Boundary conditions } \eqref{eq:boundary_condition}
\end{split}\label{eq:optimization_problem}
\end{equation}
where $k\in 0\dots N-1$ is the discretization node index, $N$ being the number of discretization nodes.
\vspace{-3mm}
\subsection{IO-linearization and zero-dynamics}
\label{sec:iolin}
\vspace{-3mm}
As in the case of \cite{ma_first_2019}, the controller we incorporate is an input-output linearization controller. Consider the system \eqref{eq:nonlinear_system} with $n$ states and $m$ inputs. We now add an output $\vm{y}_a(\vm{x}) = \vm{h}(\vm{x})$ of dimension $n_y$, which we denote the actual output. We then define $\vm{y}_d(t, \vm{\alpha})$, a time-varying vector of desired outputs. Following \citet{ma_first_2019}, these are described as $4^{\textnormal{th}}$-order Bézier curves, parametrized by $\vm{\alpha}$. Lastly, we define $\vm{y}(t, \vm{x}, \vm{\alpha}) = \vm{y}_a(\vm{x}) - \vm{y}_d(t, \vm{\alpha})$ so that $\vm{y}(t, \vm{x}, \vm{\alpha}) = 0$ whenever $\vm{y}_a(\vm{x}) = \vm{y}_d(t, \vm{\alpha})$. The behavior of $\vm{y}(t, \vm{x}, \vm{\alpha})$ is referred to as the output-dynamics, while the behavior of the system when $\vm{y}(t, \vm{x}, \vm{\alpha}) \equiv 0$ is denoted the zero-dynamics.

As described in \citet{henson_nonlinear_1997}, if the system has a well-defined \textit{vector relative degree} $\left\{r_i\right\}_{i \in 1\dots n_y}$, it may be transformed into an equivalent system where a subset of the states are the outputs of the original system along with their derivatives up to $r_i$. We may now define $\vm{y}_r \triangleq \left(y_1^{(r_1-1)}, \dots, y_{n_y}^{(r_{n_y}-1)}\right)$. In order to achieve the linear output-dynamics $\dot{\vm{y}}_r = \vm{v}$, given the assumption of well-defined relative degree, we choose the input $\vm{u}$ to be
\begin{subequations}
\begin{align}
    \!\!\vm{u}(t, \vm{x}, \vm{\alpha}) &= \mathcal{A}^{-1}(\vm{x})\left(-\vm{b}(\vm{x}) + \vm{y}_{d,r}(t, \vm{\alpha}) + \vm{v}\right)\label{eq:IO_linearization}\\
    \!\!&\textnormal{where} \nonumber\\
    \!\!\vm{b}(\vm{x}) &= \left(L^{r_1}_{\vm{f}}h_1(\vm{x}), \dots, L^{r_{n_y}}_{\vm{f}}h_{n_y}(\vm{x})\right)^\top\\
    \!\!\vm{y}_{d,r}(t, \!\vm{\alpha}) \!&= \left(y_{d,1}^{(r_1)}(t, \vm{\alpha}), \dots, y_{d,i}^{(r_i)}(t, \vm{\alpha})\right)^{\top}\\
    \!\!\mathcal{A}(\vm{x}) &=\!\!\!
    \begin{bmatrix}
        L_{\vm{g}_1}L^{r_1-1}_{\vm{f}}h_1(\vm{x}) & \dots & L_{\vm{g}_m}L^{r_1-1}_{\vm{f}}h_1(\vm{x})\\
        \vdots & \ddots & \vdots\\
        L_{\vm{g}_1}\!L^{r_{n_y}-1}_{\vm{f}}\!h_{n_y}(\vm{x}) &\!\! \dots &\!\! L_{\vm{g}_m}\!L^{r_{n_y}-1}_{\vm{f}}\!h_{n_y}(\vm{x})
    \end{bmatrix}\label{eq:general_decoupling_matrix}
\end{align}
\end{subequations}
We refer to $\mathcal{A}(\vm{x})$ as the decoupling matrix. The full sets of equations can be found in e.g. \citet[pp.~160-164]{henson_nonlinear_1997}, although we here express the output-dynamics in original coordinates in the interest of brevity.

We opt for an ambling gait where two legs are in contact with the ground at any point, and as such the system has 11 actuated \glspl{dof} (all leg joints with the constraint of constant distance between stance legs). The method requires an equal number of outputs and actuated \glspl{dof} (\cite{hereid_dynamic_2018}). In the interest of a most direct comparison with the method of \cite{ma_first_2019}, we choose the leg joints except the rear stance hip pitch as outputs.
\vspace{-3mm}
\subsection{Exponential Orbital Stabilization of full state}
\label{sec:exponential_stabilization}
\vspace{-3mm}
Although the IO-linearization controller stabilizes the output-dynamics exponentially, the system as a whole may still be unstable. It is desirable that the closed-loop system has an exponentially stable orbit in the state modulo the horizontal position, as the robot should move forward. The stability analysis of an orbit can be performed by considering the state trajectory at intersections with a transverse surface as a discrete dynamical system, described by a Poincar\'e map. The orbit will be \gls{eos} when the spectral radius of the linearization of the map around the point of intersection is less than 1 \citep{akbari_hamed_exponentially_2015}.

As in \cite{ma_first_2019}, we now consider the outputs $\vm{y}$ in the optimized controller as linear combinations of system states and corresponding optimal trajectories, parametrized by weights $\vm{\theta}$, i.e. $\vm{y}(t, \vm{x}, \vm{\alpha}^{*}, \vm{\theta})$. The outputs chosen during gait generation gives our initial guess $\vm{\theta}_0$. We then formulate the problem of exponential orbital stabilization as a \gls{bmi}-constrained optimization problem in $\vm{\theta}$. Solving this optimization problem leaves us with a set of weights $\vm{\theta}^{*}$ determining $\vm{y}(t, \vm{x}, \vm{\alpha}^{*}, \vm{\theta}^{*})$ which makes the resulting gait \gls{eos} (modulo horizontal position). See \cite{hamed_dynamically_2019,akbari_hamed_exponentially_2015} for further details and proof of the resulting stability.
\vspace{-3mm}
\subsection{Over-actuated input-output linearization}
\label{sec:overconstrained_iolin}
\vspace{-3mm}
In order for $\mathcal{A}(\vm{x})$ in \cref{eq:IO_linearization} to be invertible we must have $m = n_y$. Now, consider the case where $m > n_y$, i.e. the robot is overactuated. ASTRo, which is a quadrupedal robot, will typically have two stance feet during some phases of a gait. During these phases the number of outputs, which must equal the number of actuated \glspl{dof}, is 11, while $m=12$. As a result, the output-dynamics of the IO-linearized system are over-actuated. In \cite{ma_first_2019} this discrepancy is resolved by disabling one of the inputs, so that $m=11$. We here suggest a different approach.

We will assume that the now non-square $\mathcal{A}(\vm{x}) \in \mathbb{R}^{n_y \times m}$ has constant rank $n_y$, which is just to say that at any point in time there is a subset of inputs which can be removed to recoup the system properties assumed in the case where $n_y=m$. We can choose the feedback-law as
\begin{equation}
    \label{eq:modified_feedback_law}
    \vm{u}(t, \vm{x}) = \mathcal{A}^{+}(\vm{x})\left(-\vm{b}(\vm{x}) + \vm{y}_{d,r}(t) + \vm{v}\right)
\end{equation}
where $(\cdot)^{+}$ denotes the \gls{mpp}, to again achieve $\dot{\vm{y}}_r = \vm{v}$ as desired.

Most importantly, \eqref{eq:modified_feedback_law} yields exactly the linear, decoupled output-dynamics in the overactuated case, as \eqref{eq:IO_linearization} does in the non-overactuated case. This is due to the system of linear equations which $\mathcal{A}^{+}$ solves being underdetermined rather than overdetermined, and thus admitting of infinitely many exact solutions.

Secondly, recall that the matrix $\mathcal{A}(\vm{x})$ is assumed to have the constant rank $n_y$. As a result, the \gls{mpp} is a continuous and continuously differentiable function of the original matrix \citep{doi:10.1137/0710036}. Thus, assuming continuity and continuous differentiability of $\mathcal{A}(\vm{x})$ with respect to $\vm{x}$, these properties also hold for $\mathcal{A}^{+}(\vm{x})$. As a result, the control law $\vm{u}(t, \vm{x})$ is continuously differentiable in $\vm{x}$ for an appropriate choice of $\vm{v}$, a typical choice being a linear combination of elements from $\vm{y}$ and their derivatives up to respective relative degree.

Finally, as the \gls{mpp} characterizes the least-squares solution to an under-determined linear system, the modified controller gives the point-wise smallest control signal that results in the desired output-dynamics in the least-squares sense. In the case where the input signal is a torque, this results in a controller which utilizes all available actuators to minimize torque expenditure along the trajectory.

The minimization in the least-squares sense is also particularly desirable for the following reason. The energy expenditure of electric actuators can be divided into mechanical energy and Joule heating. The Joule heating, which was in the case of \cite{seok_design_2015} found to account for 75\% of expended energy, is well approximated as proportional to the square of applied torque. Thus, it is expected that minimizing the applied torque in a least-squares sense will reduce the expended energy as a result.

It should be noted here that the identical closed-loop system behavior in general holds only with respect to the output-dynamics; the effects on the unobservable states are not discussed. However, the design of the controller in the conventional case is performed only so as to exponentially stabilize the output-dynamics, and the stability of the zero-dynamics is subject to further analysis in any case. Just as in the case of \citet{ma_first_2019}, the post-processing described in \cref{sec:exponential_stabilization} ensures that the dynamics are \gls{eos} along the zero-dynamics manifold.
\vspace{-3mm}
\section{Results}
\label{sec:results}
\vspace{-3mm}
We now present the results with regards to energy and torque expenditure, where we compare between the method as described in \cite{ma_first_2019} and the method with our modified controller as described in \cref{sec:overconstrained_iolin}.

\begin{figure}
    \centering
    \includegraphics[width=0.28\textwidth]{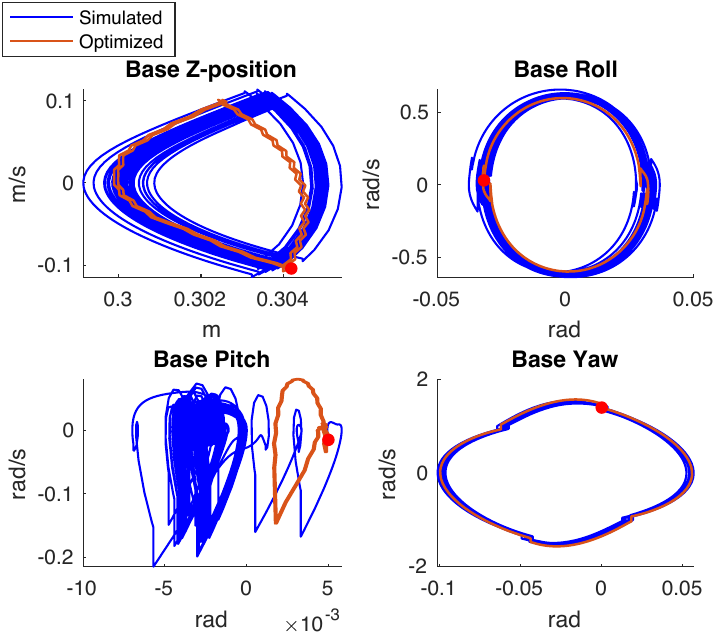}
    \caption{Phase portraits for base of optimized and simulated ambling gait. Red dots signify initial values.}
    \label{fig:ASTRo_base_phase_portrait}
\end{figure}
\begin{figure}
    \centering
    \includegraphics[width=0.46\textwidth]{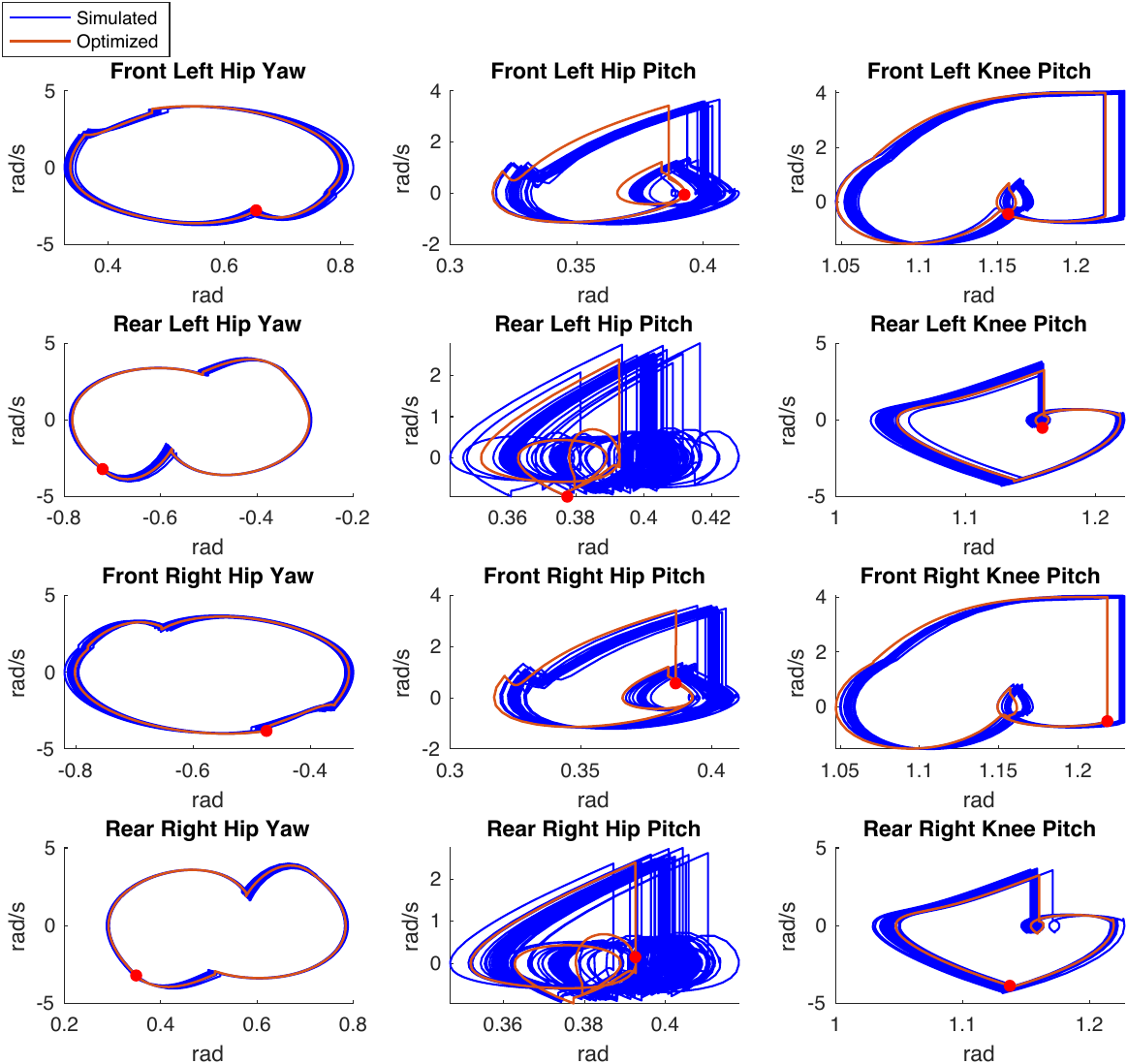}
    \caption{Phase portraits for joints of optimized and simulated ambling gait. Red dots signify initial values.}
    \label{fig:ASTRo_joint_phase_portrait}
\end{figure}

\begin{figure}
    \centering
    \includegraphics[trim=30 30 20 20,clip,width=0.38\textwidth]{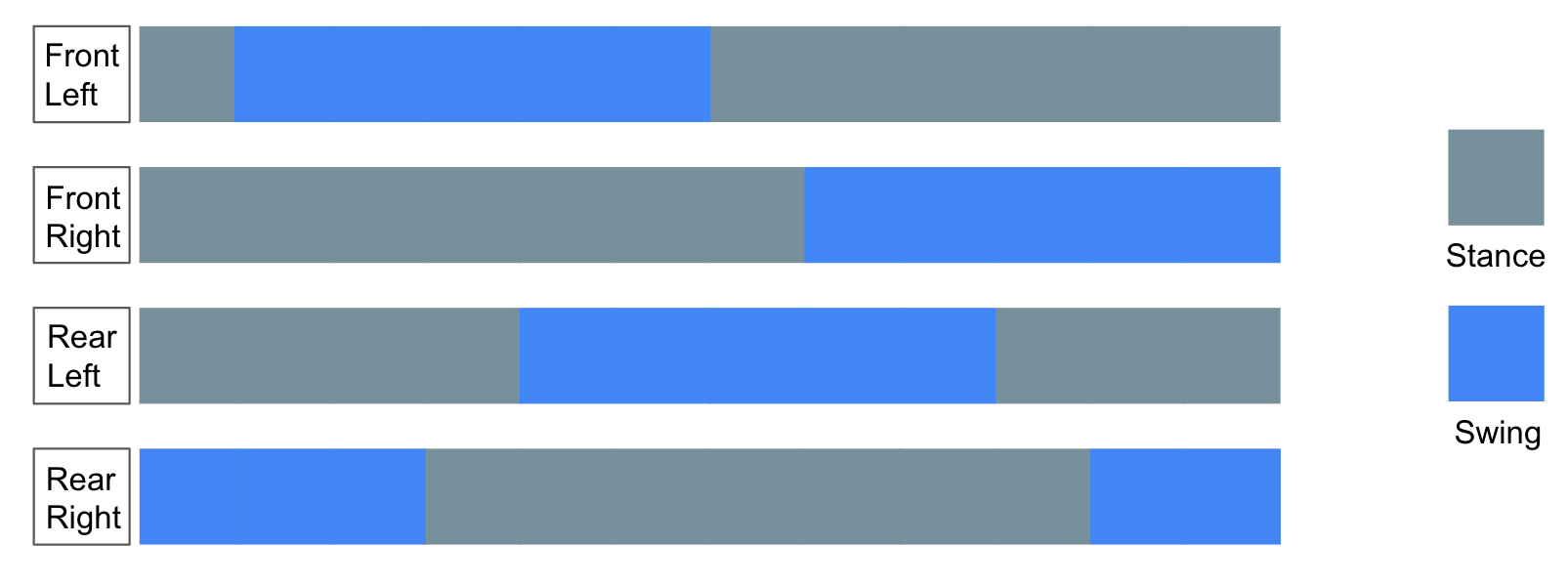}
    \caption{Diagram for one cycle of an ambling gait.}
    \label{fig:ambling_diagram}
\end{figure}

We solve the optimization problem posed in \cref{eq:optimization_problem}, using $\SI{0.2}{\meter\per\second}$ as the forward velocity in the boundary condition, and choosing the \gls{hds} graph to produce an ambling gait seen in the diagram in \cref{fig:ambling_diagram}. The controller in \cref{eq:modified_feedback_law} is used during optimization. A post-processing of the outputs is performed as described in \cref{sec:exponential_stabilization}. During this procedure the spectral radius of the linearized Poincar\'e map of the orbit goes from 1.2024 to 0.6267, rendering the orbit \gls{eos}. The resulting set of parameters $\vm{\alpha}^{*}$ and $\vm{\theta}^{*}$ define our outputs $\vm{y}$ for the controller. The gait is then simulated for 500 gait cycles using FROST. The phase portraits for the optimized and simulated state trajectories are shown in \cref{fig:ASTRo_base_phase_portrait,fig:ASTRo_joint_phase_portrait}.  Trajectories for the $x$- and $y$-positions do (by design) not constitute orbits and are omitted.

The system is simulated using both the method proposed by \cite{ma_first_2019}, where 11 actuators are utilized, and with the controller \cref{eq:modified_feedback_law} proposed in \cref{sec:overconstrained_iolin}, allowing 12 actuators to be utilized. In \cref{tab:performance} the peak and \gls{rms} torques, as well as the \gls{cot} are reported. The \gls{cot} is defined as $\frac{E}{mgd}$ where $E$ is the expended energy, $m$ and $g$ are the mass of the robot and the gravitational constant and $d$ is the total distance traveled. The following choices for modeling energy expenditure are made: As in \cite{seok_design_2015}, we calculate the power expenditure of each actuator as the sum of expended mechanical power and power lost to heating effects. For the mechanical energy, we choose a simple model with no recuperation of absorbed mechanical energy through e.g. regenerative braking. We thus calculate expended mechanical power as $P_{\textnormal{mec}} = \max{(\tau\times\omega, 0)}$ for each motor, where $\tau$ is the joint torque and $\omega$ is the angular velocity. The power lost to heating effects as a function of the joint torque is written as $P_{\textnormal{heat}} = R\left(\frac{\tau_{\textnormal{motor}}}{K_t}\right)^2$, where $K_t$ is the motor torque constant, $R$ is the resistance of the motor and $\tau_{\textnormal{motor}}$ is the motor torque. The motor torque is defined as $\tau_{\textnormal{motor}} = \frac{\tau}{N}$ where $N$ is the gear ratio between the motor and the joint. For calculating heating effects as a function of joint torques, we require properties of a specific actuator which is not a part of the model used in simulation. In the interest of performing these calculations with realistic parameter values, we base ourselves on the RMD-X8-Pro-H from MyActuator, with an additional gear ratio of 6:1, i.e. $N=6$. The actuator has a $K_t = \SI{0.29}{\newton\meter\per\ampere}$ and an $R = \SI{0.55}{\ohm}$. The gear ratio is chosen to bring the torque trajectories within the nominal torque operating range of the actuator while retaining the resulting angular velocities within the geared angular velocity operating range. The \gls{cot} can then be written
\begin{equation}
    \textnormal{CoT} = \frac{1}{mgd}\int_{t_0}^{t_f}\sum_{\textnormal{actuators}}\left(P_{\textnormal{mec}}(t) + P_{\textnormal{heat}}(t)\right)\mathrm{d}t
\end{equation}

\begin{table}[]
    \centering
    \begin{tabular}{l|c|c|c}
    \hline
    \# Method & \gls{cot} & Peak torque & Torque \gls{rms}\\
    \hline
    \cite{ma_first_2019} & 6.389 & \SI{83.97}{\newton\meter} & \SI{9.659}{\newton\meter}\\
    Proposed & 4.532 & \SI{66.55}{\newton\meter} & \SI{7.932}{\newton\meter}\\
    \end{tabular}
    \caption{Performance metrics for previous and proposed method}
    \label{tab:performance}
\end{table}

As we can see in \cref{tab:performance}, our proposed method yields a fairly substantial reduction in both the \gls{cot}, peak torque and \gls{rms} torque. The \gls{cot} is reduced by $\approx{29.1}\%$. The \gls{rms} torque is reduced by $\approx{17.9}\%$, and the peak torque is reduced by $\approx{20.7\%}$. This indicates that the modification, in addition to avoiding arbitrary choices of which inputs to disable, may actually lead to fairly large energy savings in practice, without at all having to modify the output trajectories of the system.

\begin{figure}
    \centering
    \includegraphics[width=0.42\textwidth]{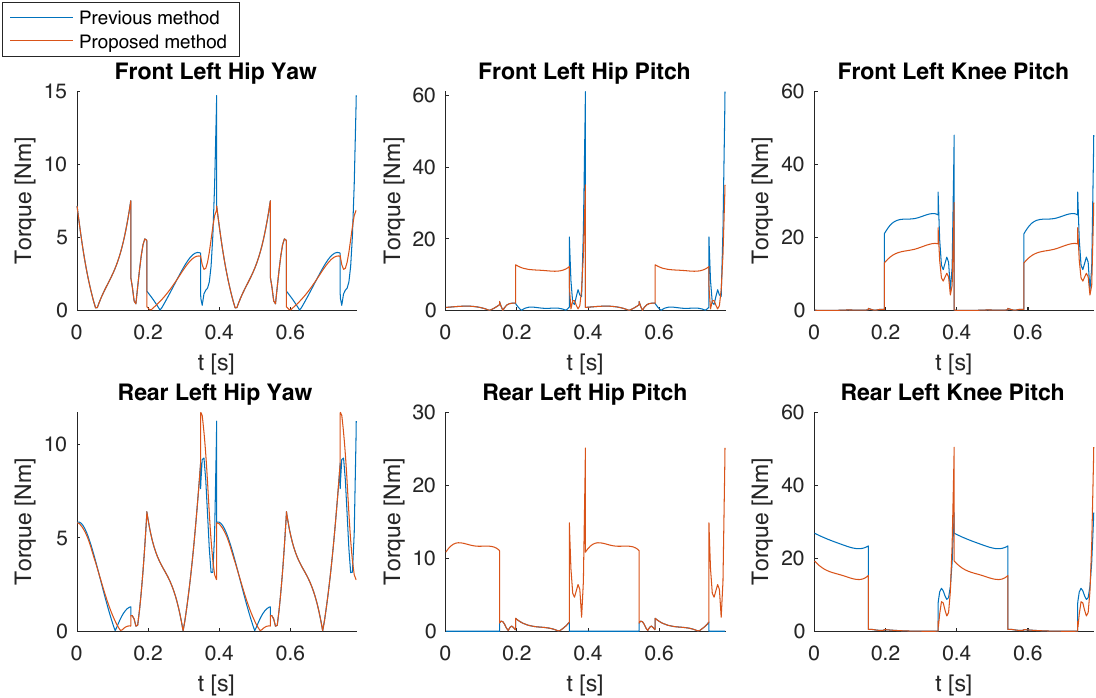}
    \caption{Plots of absolute value of torques for left leg actuators for two consecutive gait cycles.}
    \label{fig:torque_plots}
\end{figure}

In \cref{fig:torque_plots}, in order to investigate the differences in the torque profiles more closely, we have plotted the absolute value of the torques of the left leg actuators for both methods. Due to the symmetry of the ambling gait, the right leg actuator torques will be identical but phase-shifted by half a gait cycle. As can be seen, the rear hip pitch torque is much greater for our method compared to the method proposed in \cite{ma_first_2019}. This is to be expected: In the original method the rear hip pitch actuators are disabled during stance phases to avoid overactuation. Thus, the rear hip pitch actuators are only active during swing leg phases, in which the actuator only lifts the leg off the ground. The front hip pitch actuator is also seen to be more active for the proposed method, although the picture is not as clear since the method from \cite{ma_first_2019} exhibits significantly higher peak torques. The increased torque use in hip pitch actuators in our method, however, is seen to lead to substantially lower torque expenditure in all knee actuators compared to the method from \cite{ma_first_2019}. This even distribution of load-bearing between actuators, along with heat-related power losses depending on the square of torque expenditure, explains the significant reduction in \gls{cot}.

\vspace{-2mm}
\section{Conclusions and future work}
\label{sec:conclusion}
\vspace{-2mm}
In this paper we have presented a modification to an existing method for controller synthesis through offline nonlinear optimal control. The modification allows the controller to better utilize all available inputs in overactuated scenarios. Simulations indicate that the savings in torque expenditure and \gls{cot} may be quite significant in certain cases when compared to the original method, without requiring any modification of the system trajectories.

Future work includes experimental validation of the torque reduction which we have here observed in simulations, as well as exploring further modifications to the controller with the aim of increasing the robustness.

%\begin{spacing}{0.9}
\bibliographystyle{ifacconf.bst}
%\bibliography{bib/references}
\bibliography{bib/references_shortened}
%\end{spacing}
%\appendix

\end{document}